# A Simple Channel Independent Beamforming Scheme With Parallel Uniform Circular Array

Haiyue Jing, *Student Member, IEEE*, Wenchi Cheng, *Senior Member, IEEE*, and Xiang-Gen Xia, *Fellow, IEEE*

***Abstract*— In this letter, we consider a uniform circular array (UCA)-based line-of-sight multiple-input-multiple-output system, where the transmit and receive UCAs are parallel but non-coaxial with each other. We propose a simple channel-independent beamforming scheme with fast symbol-wise maximum likelihood detection.**

## I. Introduction

DURING the past two decades, multiple-input-multiple-output (MIMO) with uniform linear array (ULA) antenna architecture for wireless communications has been widely studied. A number of efficient de/modulation schemes have been proposed in the literature, such as, Bell Laboratories Layered Space-Time (BLAST) [1], [2] and space-time block coded (STBC) schemes [3]. The demodulation methods include linear receivers, successful interference cancelation (SIC) receivers, and maximum-likelihood (ML) receivers. Although linear and SIC receivers are usually much faster than the ML receivers, their complexities are either at least in the third order of the signal block length or/and they have poor performances. To have a fast ML receiver, orthogonal STBC (OSTBC) etc. usually need to be equipped across several time slots. However, OSTBC may lead to a multiplexing loss [4].

Recently, uniform circular arrays (UCAs) have been recognized as a useful alternative for wireless communications over line-of-sight (LOS) channels [5]. The channel matrix is a circulant matrix when the transmit and the receive UCAs are aligned with each other [5], [6]. The LOS transmission is a very important scenario, such as LOS millimeter wave communications [7] and terahertz communications [8]. Chen *et al.* [9] investigated the impact of misalignment on the capacity. They focused on the scenario where the projection of the center of the receive UCA along the $z$-axis of the transmit UCA is on the coordinate axis $x$ or $y$ of the transmit UCA, instead of a generic case. In this letter, we propose a simple channel independent beamforming scheme for the UCA based LOS MIMO system with the fast symbol-wise ML detection in the parallel but non-coaxial case. Our proposed beamforming scheme first converts the channel from a non-circulant matrix to a circulant matrix, and then it becomes equivalent to the coaxial case, i.e., the transmit and receive UCAs are aligned. After that, it is similar to the orthogonal frequency-division multiplexing (OFDM) system and instead of transmitting $N$ subcarrier signals across $N$ time slots in an OFDM system, it transmits them across $N$ transmit antennas in one time slot. It requires that $N$ receive antennas are equipped as well. When the channel matrix is converted to a circulant matrix, the discrete Fourier transform (DFT)/inverse DFT (IDFT) can diagonalize the channel matrix similar to that for the intersymbol-interference (ISI) channel matrix after the cyclic prefix (CP) inserted/deleted in an OFDM system. After the diagonalization, the fast symbol-wise ML detection holds. We want to emphasize that our proposed scheme has symbol rate $N$, i.e., $N$ symbols per channel use. We then derive the bit error rate (BER) for the LOS MIMO system. Also, we compare the BER and the number of computations for the LOS MIMO system using the channel independent beamforming with those of the traditional LOS MIMO system. Simulation results verify that our proposed channel independent beamforming scheme with the fast symbol-wise ML detection has the same BER performance as the traditional scheme, while the computational complexity is much lower.

## II. System Model for UCA Based MIMO Communications

Figure 1 shows the system model for LOS MIMO based wireless communications, where the antennas of transmit and receive UCAs are uniformly around the perimeter of the circle. In this model, the transmit and receive UCAs are parallel to each other. We denote by $r$ and $R$ the radii of the transmit and receive UCAs, respectively, where $r$ and $R$ can be different. The notation $d$ represents the distance between the center of transmit UCA and the center of receive UCA while $d_{mn}$ represents the distance from the $n$th transmit antenna to the $m$th receive antenna. The notations $\widetilde{d}_r$ and $\widetilde{d}_R$ represent the distances between two neighboring antennas of the transmit and receive UCAs, respectively. We denote by $\alpha_r$ and $\alpha_R$ the angles between the phase angle of the first antenna and zero radian corresponding to the transmit and receive UCAs, respectively. There are $N$ and $M$ antennas at the transmit and receive UCAs, respectively. The parameter $\theta$ denotes the included angle between $x$-axis and the projection of the line

Manuscript received December 4, 2018; accepted December 22, 2018. Date of publication January 11, 2019; date of current version March 8, 2019. The associate editor coordinating the review of this paper and approving it for publication was C.-K. W. Wen. This work was supported in part by the National Natural Science Foundation of China (No. 61771368), the Young Elite Scientists Sponsorship Program by CAST (2016QNRC001), and the Young Talent Support Fund of Science and Technology of Shaanxi Province (2018KJXX-025). *(Corresponding author: Wenchi Cheng.)*

H. Jing and W. Cheng are with the State Key Laboratory of Integrated Services Networks, Xidian University, Xi'an 710071, China (e-mail: hyjing@stu.xidian.edu.cn; wccheng@xidian.edu.cn).

X.-G. Xia is with Xidian University, Xi'an 710071, China, and also with the Department of Electrical and Computer Engineering, University of Delaware, Newark, DE 19716 USA (e-mail: xxia@ee.udel.edu).

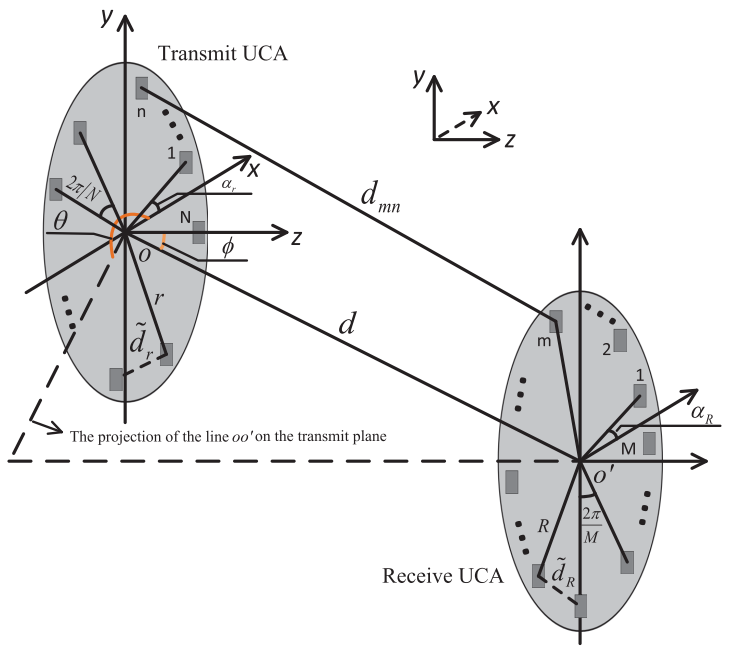

Fig. 1. The system model for LOS MIMO based wireless communications.

from the center of the transmit UCA ($o$) to the center of the receive UCA ($o'$) on the transmit plane. Also, $\phi$ denotes the included angle between $z$-axis and the line from the center of the transmit UCA to the center of the receive UCA. For the LOS scenario as studied in this letter, it is easy for the transmitter and receiver to know $\theta$ and $\phi$. For the channel independent beamforming scheme, the antennas are fed with the same input signal but with different phase factors. Our proposed channel independent beamforming scheme will be presented in the next section. Note that the antennas are fed with different signals for the traditional LOS MIMO based wireless communications.

## III. Channel Model for UCA Based LOS MIMO

We denote by $h_{m,n}$ the channel gain from the $n$th antenna on the transmit UCA to the $m$th antenna on the receive UCA. Then, $h_{mn}$ can be written as follows [10]:

$$h_{m,n} = \frac{\beta\lambda e^{-j\frac{2\pi}{\lambda}d_{mn}}}{4\pi d_{mn}}, \tag{1}$$

where $\beta$ denotes the combination of all the relevant constants, such as, attenuation and phase rotation caused by antennas and their patterns on both sides.

Let the center of the transmit UCA as the origin of the coordinate system and the transmit UCA is on the $x-y$ plane. The coordinate of the center corresponding to the receive UCA is $(d\sin\phi\cos\theta, d\sin\phi\sin\theta, d\cos\phi)$. Thus, the coordinate of the $m$th receive antenna, denoted by $(B_x, B_y, B_z)$, is $(d\sin\phi\cos\theta + R\cos(\psi_m+\alpha_R), d\sin\phi\sin\theta + R\sin(\psi_m + \alpha_R), d\cos\phi)$, where $\psi_m + \alpha_R = 2\pi(m-1)/M + \alpha_R$ is the phase angle of the $m$th receive antenna. Then, $d_{mn}$ is derived as follows:

$$\begin{aligned} d_{mn} &= \sqrt{[B_x - r\cos(\varphi_n+\alpha_r)]^2 + [B_y - r\sin(\varphi_n+\alpha_r)]^2 + B_z^2} \\ &= \sqrt{d^2 + R^2 + r^2 + 2\mathcal{E}_m + 2\mathcal{D}_{mn} + 2\mathcal{F}_n} \end{aligned} \tag{2}$$

where $\varphi_n + \alpha_r = 2\pi(n-1)/N + \alpha_r$ is the phase angle of the $n$th transmit antenna, and $\mathcal{E}_m$, $\mathcal{D}_{mn}$, and $\mathcal{F}_n$ are given by

$$\begin{cases} \mathcal{E}_m = dR\sin\phi\cos(\psi_m + \alpha_R - \theta); \\ \mathcal{D}_{mn} = -Rr\cos(\varphi_n - \psi_m + \alpha_r - \alpha_R); \\ \mathcal{F}_n = -dr\sin\phi\cos(\varphi_n + \alpha_r - \theta). \end{cases} \tag{3}$$

Since $d \gg r$ and $d \gg R$, we can make approximations for $d_{mn}$ at the denominator and numerator of Eq. (1). For the denominator, we may simply use the approximation $d_{mn} \approx d$ since it is insignificant compared to that of the numerator. For the numerator, $d_{mn}$ can be approximated according to $\sqrt{1+2x} \approx 1 + x$ when $x$ is close to zero as follows:

$$d_{mn} \approx \sqrt{d^2 + R^2 + r^2} + \frac{\mathcal{E}_m + \mathcal{D}_{mn} + \mathcal{F}_n}{\sqrt{d^2 + R^2 + r^2}}. \tag{4}$$

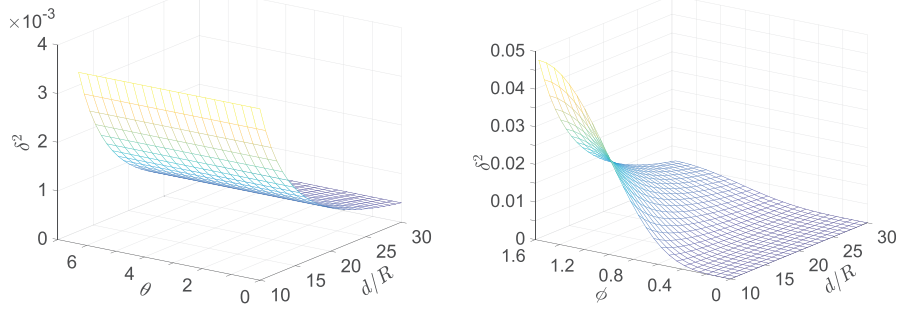

(a) $\delta^2$ versus $d/R$ and $\theta$. (b) $\delta^2$ versus $d/R$ and $\phi$.

Fig. 2. The parameter $\delta^2$ versus different parameters.

Substituting Eq. (4) into Eq. (1), we can obtain the approximated channel gain, denoted by $\widehat{h}_{m,n}$, as follows:

$$\begin{aligned} \widehat{h}_{m,n} = \frac{\beta\lambda}{4\pi d}\exp\left\{-j\frac{2\pi}{\lambda}\sqrt{d^2 + r^2 + R^2}\right\} \\ \times \exp\left\{-j\frac{2\pi(\mathcal{E}_m + \mathcal{D}_{mn} + \mathcal{F}_n)}{\lambda\sqrt{d^2 + r^2 + R^2}}\right\}. \end{aligned} \tag{5}$$

To evaluate the difference between $h_{m,n}$ and $\widehat{h}_{m,n}$, we denote by $\delta^2$ the variance of the difference as follows:

$$\delta^2 = \frac{\sum_{n=1}^{N}\sum_{m=1}^{M}|h_{m,n} - \widehat{h}_{m,n}|^2}{\sum_{n=1}^{N}\sum_{m=1}^{M}|h_{m,n}|^2}. \tag{6}$$

Figure 2 plots the variance $\delta^2$ versus different parameters, where $\beta = 4\pi$, $\lambda = 0.1$ m, $R = r = 0.4$ m, $\alpha_r = \alpha_R = 0$, and $M = N = 10$. As shown in Fig. 2, $\theta$ has little impact on the variance. More importantly, we can observe that $\delta^2$ is close to zero when $d \gg R$. Thus, the approximations have little impact on the channel gain and we can use Eq. (5) to approximate Eq. (1) for the following derivations.

If the number of antennas at the transmitter is equal to the number of antennas at the receiver, i.e., $M = N$, the channel matrix $\boldsymbol{H}$ can be written as follows:

$$\boldsymbol{H} = \boldsymbol{\Delta}\widetilde{\boldsymbol{H}}\boldsymbol{\Gamma} \tag{7}$$

where $\boldsymbol{\Delta}$ and $\boldsymbol{\Gamma}$, given by

$$\begin{cases} \boldsymbol{\Delta} = \mathrm{diag}\left(e^{-j\frac{2\pi\mathcal{E}_1}{\lambda\sqrt{d^2+r^2+R^2}}}, \cdots, e^{-j\frac{2\pi\mathcal{E}_N}{\lambda\sqrt{d^2+r^2+R^2}}}\right), \\ \boldsymbol{\Gamma} = \mathrm{diag}\left(e^{-j\frac{2\pi\mathcal{F}_1}{\lambda\sqrt{d^2+r^2+R^2}}}, \cdots, e^{-j\frac{2\pi\mathcal{F}_N}{\lambda\sqrt{d^2+r^2+R^2}}}\right), \end{cases} \tag{8}$$

are diagonal matrices with unit norm diagonal elements, thus unitary. The matrix $\widetilde{\boldsymbol{H}}$ is with elements

$$\widetilde{h}_{m,n} = \frac{\beta\lambda}{4\pi d}\exp\left\{\frac{-j2\pi}{\lambda}\left(\sqrt{d^2+r^2+R^2} + \frac{\mathcal{D}_{mn}}{\sqrt{d^2+r^2+R^2}}\right)\right\}, \tag{9}$$

where $1 \leq n, m \leq N$. Since the matrix $\mathbf{D} = (\mathcal{D}_{mn})$, as shown in Eq. (3), is a circulant matrix. Therefore, the matrix $\widetilde{\boldsymbol{H}}$ is also a circulant matrix. Since $\widetilde{\boldsymbol{H}}$ is a circulant matrix, the DFT matrix can diagonalize it:

$$\boldsymbol{W}^*\widetilde{\boldsymbol{H}}\boldsymbol{W} = \boldsymbol{\Lambda}, \tag{10}$$

where $()^*$ denotes the conjugate transpose,

$$\boldsymbol{\Lambda} = \mathrm{diag}\left(H_{1,1}, H_{1,2}, \cdots, H_{1,N}\right), \tag{11}$$

and $\boldsymbol{W}$ is the standard $N \times N$ IDFT matrix with elements $\left\{\frac{1}{\sqrt{N}}\exp[j2\pi(n-1)(l-1)/N]\right\}$, $1 \le l, n \le N$. In Eq. (11), the sequence $\{H_{1,1}, H_{1,2}, \cdots, H_{1,N}\}$ is the $N$-point DFT of the sequence $\left\{\widetilde{h}_{1,1}, \widetilde{h}_{1,2}, \cdots, \widetilde{h}_{1,N}\right\}$ corresponding to the first row of the matrix $\widetilde{\boldsymbol{H}}$, i.e.,

$$H_{1,k} = \sum_{n=1}^{N} \widetilde{h}_{1,n} e^{-j\frac{2\pi(n-1)(k-1)}{N}}. \tag{12}$$

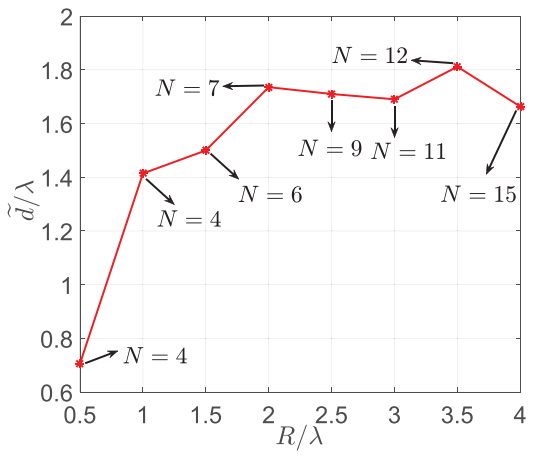

Fig. 3. The distance between two neighboring antennas with $\sigma^2 < 0.01$.

Since $H_{1,k}$ are just the $N$-point DFT of the original channel coefficients $\widetilde{h}_{1,n}$ and the DFT is unitary, the capacity does not change.

The variance, denoted by $\sigma^2$, of $\{|H_{1,1}|, |H_{1,2}|, \cdots, |H_{1,N}|\}$ is given as follows:

$$\sigma^2 = \frac{1}{N}\sum_{k=1}^{N}\left(|H_{1,k}| - \frac{1}{N}\sum_{i=1}^{N}|H_{1,i}|\right)^2. \tag{13}$$

Similar to the conventional OFDM system, when the variance of $|H_{1,k}|$ along $k$ is smaller, i.e., $\sigma^2$ is smaller, the channel is better.

Fig. 3 gives the normalized distance by wavelength, i.e., $\widetilde{d}/\lambda$, between two neighboring antennas with $\sigma^2 < 0.01$ for the special scenario where the transmit and receive UCAs are aligned with each other. In Fig. 3, $\widetilde{d} = \widetilde{d}_r = \widetilde{d}_R$. As $R$ increases, the maximum possible number of antennas increases given $\sigma^2 < 0.01$. $\widetilde{d}$ is roughly $1.7\lambda$ when $R$ is larger than $2\lambda$, corresponding to the optimal $N$ with $\sigma^2 < 0.01$. Thus, we can follow this setup to avoid that the channels are too correlated.

## IV. Achieving Fast Detection for UCA Based LOS MIMO Communications

### A. Channel Independent Beamforming Scheme

We propose a channel independent beamforming scheme, where each information symbol is transmitted using all antennas with different phase factors on each antenna. The power of each information symbol is averagely allocated to $N$ transmit antennas. Let $\{s_1, s_2, \cdots, s_N\}$ be $N$ information symbols to be sent and $x_n$ $(n = 1, 2, \cdots, N)$ be the symbol to be transmitted at the $n$th transmit antenna. Then, we propose to use the following modulation:

$$x_n = \sum_{l=1}^{N}\frac{1}{\sqrt{N}} s_l e^{j\frac{2\pi(n-1)}{N}(l-1)} e^{j\frac{2\pi\mathcal{F}_n}{\lambda\sqrt{d^2+r^2+R^2}}}. \tag{14}$$

According to Eq. (14), we can write the vector, denoted by $\boldsymbol{x}$, corresponding to the transmit signals at the transmit UCA as follows:

$$\boldsymbol{x} = \boldsymbol{\Gamma}^*\boldsymbol{W}\boldsymbol{s}, \tag{15}$$

where $\boldsymbol{s} = [s_1, s_2, \cdots, s_N]^T$ is the information symbol vector. When $\mathcal{F}_n = 0$, i.e., the transmit and receive UCAs are coaxial with each other, the expression of the transmit signal is similar to that of the traditional orthogonal frequency-division multiplexing (OFDM) [11]. But it is different in the sense that the transmission here is along the antenna direction, i.e., the $n$th subcarrier signal $x_n$ is transmitted in the $n$th transmit antenna instead of the $n$th time slot. In addition, all the $N$ subcarrier symbols are transmitted in one time slot, which means that the symbol rate is $N$, i.e., $N$ symbols per channel use. Also, different from OFDM, the above proposed scheme in Eq. (14) does not add any CP in the transmission. Note that this case, i.e., coaxial case, has been studied in [5] and [6].

### B. Fast Symbol-Wise ML Detection Scheme

The received signal vector for the LOS MIMO system, denoted by $\boldsymbol{y}$, can be derived as follows:

$$\boldsymbol{y} = \boldsymbol{H}\boldsymbol{x} + \boldsymbol{z} = \boldsymbol{H}\boldsymbol{\Gamma}^*\boldsymbol{W}\boldsymbol{s} + \boldsymbol{z}, \tag{16}$$

where $\boldsymbol{y} = [y_1, \cdots, y_m, \cdots, y_N]^T$ and $\boldsymbol{z} = [z_1, \cdots, z_m, \cdots, z_N]^T$ with $y_m$ and $z_m$ representing the receive signal and the noise of the $m$th receive antenna.

To obtain the transmitted information symbols, we multiply the matrix $\boldsymbol{W}^*\boldsymbol{\Delta}^*$ to the received signal vector $\boldsymbol{y}$:

$$\boldsymbol{W}^*\boldsymbol{\Delta}^*\boldsymbol{y} = \boldsymbol{W}^*\boldsymbol{\Delta}^*\boldsymbol{\Delta}\boldsymbol{W}\boldsymbol{\Lambda}\boldsymbol{s} + \boldsymbol{W}^*\boldsymbol{\Delta}^*\boldsymbol{z} = \boldsymbol{\Lambda}\boldsymbol{s} + \boldsymbol{W}^*\boldsymbol{\Delta}^*\boldsymbol{z}. \tag{17}$$

Since the product matrix $\boldsymbol{W}^*\boldsymbol{\Delta}^*$ is unitary, the noise statistics does not change. Denote $\widetilde{\boldsymbol{y}} = \boldsymbol{W}^*\boldsymbol{\Delta}^*\boldsymbol{y} = [\widetilde{y}_1, \widetilde{y}_2, \cdots, \widetilde{y}_N]^T$. Then, we can obtain an estimate, denoted by $\widehat{\boldsymbol{s}}$, of the transmitted information symbol vector $\boldsymbol{s}$ via the ML detection as follows:

$$\begin{aligned}\widehat{\boldsymbol{s}} &= \arg\min_{\boldsymbol{s}\in\boldsymbol{\Omega}^N}\|\widetilde{\boldsymbol{y}} - \boldsymbol{\Lambda}\boldsymbol{s}\|^2 = \arg\min_{\boldsymbol{s}\in\boldsymbol{\Omega}^N}\sum_{l=1}^{N}|\widetilde{y}_l - H_{1,l}s_l|^2 \\ &= \left[\arg\min_{s_1\in\boldsymbol{\Omega}}|\widetilde{y}_1 - H_{1,1}s_1|, \cdots, \arg\min_{s_N\in\boldsymbol{\Omega}}|\widetilde{y}_N - H_{1,N}s_N|\right]^T,\end{aligned} \tag{18}$$

where $\boldsymbol{\Omega}$ is a signal constellation of size $K$. The last equality in Eq. (18) is for an uncoded system where the information symbols $[s_1, ..., s_N]$ are independent.

For the traditional LOS MIMO system with $N$ transmit antennas, each antenna transmits one symbol. Thus, the corresponding traditional ML detection scheme is given as follows:

$$\widehat{\boldsymbol{s}} = \arg\min_{\boldsymbol{s}\in\boldsymbol{\Omega}^N}\|\boldsymbol{y} - \boldsymbol{H}\boldsymbol{s}\|^2. \tag{19}$$

TABLE I
COMPLEXITY COMPARISON

| Detection scheme | The number of complex additions | The number of complex multiplications |
|---|---|---|
| Fast symbol-wise ML | $N\log_2(N)+NK$ | $\frac{N}{2}\log_2(N)+N(K+1)$ |
| Traditional ML | $N^2K^N$ | $(N^2+N)K^N$ |

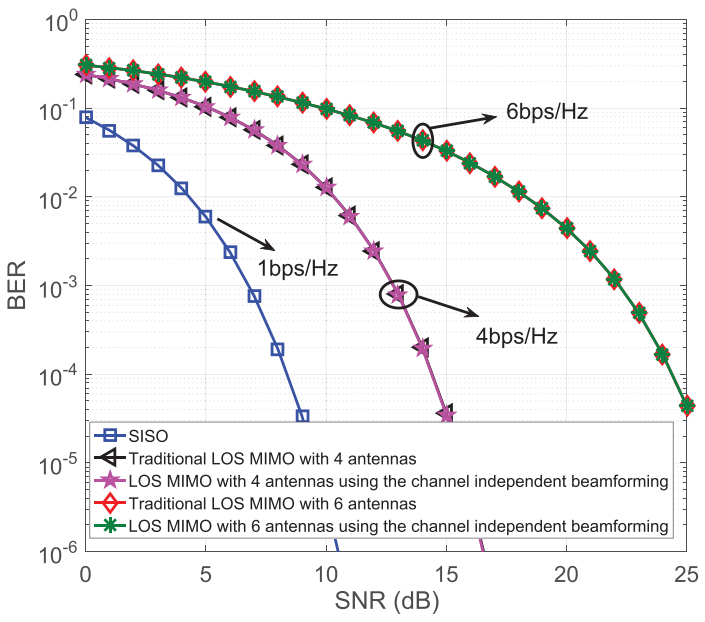

Fig. 4. BERs for the LOS MIMO system using the channel independent beamforming and the traditional LOS MIMO system (BPSK).

Note that our proposed channel independent beamforming is a basic modulation for UCA based LOS MIMO. Similar to the conventional OFDM for ISI channels, if the channel state information is known at the transmitter, further precoding can be employed across the subcarriers to improve the performance. This is particularly important when $N$ is large and the channels become too correlated.

### C. BER and Computational Complexity Analyses

The BER, denoted by $P_e$, corresponding to binary phase shift keying (BPSK) for the LOS MIMO system using the channel independent beamforming can be easily derived from Eq. (18) as follows [12]:

$$P_e=\frac{1}{N}\sum_{l=1}^{N}\frac{1}{2}\mathrm{erfc}\left(\frac{|H_{1,l}|^2|s_l|^2}{N\omega^2}\right), \tag{20}$$

where $\mathrm{erfc}(x)=\frac{2}{\sqrt{\pi}}\int_x^{\infty}e^{-t^2}dt$ and $\omega^2$ represents the variance of received noise.

The computational complexities, which are listed in Table I, of our proposed scheme with the fast symbol-wise ML detection and the traditional scheme with the ML detection are calculated by Eqs. (17), (18), and (19). In Table I, $K$ represents the size of modulation alphabet. We can observe that the numbers of complex additions and complex multiplications corresponding to the fast symbol-wise ML detection are much smaller than those corresponding to the traditional ML detection, respectively. For example, when $K=4$ and $N=8$, the numbers of additions and multiplications for the traditional ML are approximately 74898 and 90742 times, respectively, more than those for the fast symbol-wise ML.

## V. NUMERICAL RESULTS

We numerically evaluate the LOS MIMO system using the channel independent beamforming. We compare the BER for the LOS MIMO system using the channel independent beamforming and the BER of the traditional LOS MIMO system, where we set $\beta=4\pi$, $\alpha_r=\alpha_R=0$, $\theta=0$, $\phi=\pi/6$, $\lambda=0.01$ m, $R=r=0.1$ m, and $d=4$ m. In this case, the channel approximation variance $\delta^2=0.02305$ when $N=M=4$ and $\delta^2=0.01403$ when $N=M=6$.

Figure 4 shows the BERs for the LOS MIMO system using the channel independent beamforming and the traditional LOS MIMO system. We can observe that the BERs for the two schemes are the same. This is because our proposed channel independent beamformers at both transmit and receive sides are unitary transforms and therefore, they do not change the channel properties. The BER increases as the number of antennas increases, while the transmission throughput increases as well. Note that, although their BER performances are the same, as it is shown in Table I, the computational complexity of our proposed scheme is much lower.

## VI. CONCLUSIONS

In this letter, we proposed a channel independent beamforming and the corresponding fast symbol-wise ML detection, where the transmit and the receive UCAs are parallel but non-coaxial with each other. We would like to emphasize that our proposed MIMO scheme has both symbol rate $N$ and the fast symbol-wise ML detection for $N$ transmit antennas. We compared BER and the number of computations corresponding to the LOS MIMO system using the channel independent beamforming and the corresponding fast symbol-wise ML detection with those of the traditional LOS MIMO system. While the BER of the LOS MIMO system using the channel independent beamforming is the same as that of the traditional LOS MIMO system, the numbers of complex additions and complex multiplications are much smaller than those of the traditional LOS MIMO detection scheme.